\def\tr{{\rm tr}}

\def\C{{\bf C}}

\def\C#1{C^{(#1)}}
\def\half{\frac{1}{2}}
\def\RR{R$\otimes$R }   
\def\NSNS{NS$\otimes$NS }

\def\spc{\;\;\;\;}

\def\modz{\left| z\right|}
\def\mody{\left| y\right|}
\def\mdz#1{\left| z_{#1}\right|}
\def\ck{\left<\right.\!\C2,k^0\!\left.\right|}
\def\bdr{\left|B\right>}
\def\e#1{e^{ik^{#1}X}}

\def\xxx#1           {{hep-th/#1}}

\def\npb#1(#2)#3     {Nucl. Phys. {\bf B#1} (#2) #3 }
\def\rep#1(#2)#3     {Phys. Rept.{\bf #1} (#2) #3 }
\def\plb#1(#2)#3     {Phys. Lett. {\bf #1B} (#2) #3 }
\def\prl#1(#2)#3     {Phys. Rev. Lett.{\bf #1} (#2) #3 }
\def\prd#1(#2)#3     {Phys. Rev. {\bf D#1} (#2) #3 }
\def\ap#1(#2)#3      {Ann. Phys. {\bf #1} (#2) #3 }
\def\rmp#1(#2)#3     {Rev. Mod. Phys. {\bf #1} (#2) #3 }
\def\cmp#1(#2)#3     {Comm. Math. Phys. {\bf #1} (#2) #3 }
\def\mpla#1(#2)#3    {Mod. Phys. Lett. {\bf A#1} (#2) #3 }
\def\ijmp#1(#2)#3    {Int. J. Mod. Phys. {\bf A#1} (#2) #3 }
\def\cqg#1(#2)#3     {Class. Quant. Grav. {\bf #1} (#2) #3 }
\def\am#1(#2)#3      {Adv. Math. {\bf #1} (#2) #3 }
\def\im#1(#2)#3      {Invent. Math. {\bf #1} (#2) #3 }
\def\jhep#1(#2)#3    {J. High Energy Phys. {\bf #1} (#2) #3 }

\def\dstyler#1       {\documentstyle{report}[#1]}
\def\dstylea#1       {\documentstyle{article}[#1]}
\def\bd              {\begin{document}}
\def\ed              {\end{document}}
\def\be              {\begin{equation}}
\def\ee              {\end{equation}}
\def\ba              {\begin{eqnarray}}
\def\ea              {\end{eqnarray}}
\def\ni              {\noindent}
\def\bb#1            {\begin{thebibliography}{#1}}
\def\eb              {\end{thebibliography}}

\def\etal {{\em et al.} }
\def\w    {\;_\wedge}
\def\ie   {{\em i.e.}}

\documentstyle{article}[12pt]
\begin{document}

\thispagestyle{empty}
\begin{flushright}
  hep-th/9812088\\
  DAMTP-1998-166
 \end{flushright}
\vskip 0.5cm
\begin{center}\LARGE
{\bf Gravitational Couplings of D-branes and O-planes}
\end{center}
\vskip 1.0cm
\begin{center}
{\large  Bogdan Stefa\'nski, jr.\footnote{E-mail  address:
{\tt B.Stefanski@damtp.cam.ac.uk}}}

\vskip 0.5 cm
{\it Department of Applied Mathematics and Theoretical Physics\\
Cambridge University, Cambridge, England}
\end{center}

\vskip 1.0cm

\begin{center}
December 1998
\end{center}

\vskip 1.0cm

\begin{abstract} An explicit calculation is performed to check all the
tangent bundle gravitational couplings of Dirichlet branes and Orientifold
planes by scattering $q$ gravitons with a $p+1$ form Ramond-Ramond
potential in the world-volume of a $D(p+2q)$-brane. The structure of the
D-brane Wess-Zumino term in the world-volume action is confirmed, while a
different O-plane Wess-Zumino action is obtained. 
\end{abstract}

\vfill
\setcounter{footnote}{0}
\def\thefootnote{\arabic{footnote}}
\newpage

\section{Introduction}

Properties of D-branes~\cite{PolRR} have provided new insights into the
structure of string theory. The effective world-volume action of a D-brane
has two terms. The Dirac-Born-Infeld (DBI) term describes the world-volume
vector potential and scalars coupling to the Neveu-Schwarz-Neveu-Schwarz
(NS$\otimes$NS)  fields. The Wess-Zumino (WZ) term gives the coupling of
the vector potential and the pullback of the curvature fields to the
Ramond-Ramond (R$\otimes$R) potentials. D-branes have no intrinsic
gravitational dynamics. However, a D-brane is affected by the
gravitational field in the background space in which it is
embedded.\footnote{Throughout this paper we consider a trivial
normal-bundle.}

The DBI part of the action, at least for abelian gauge fields, has been
known for some time~\cite{Leigh}. The WZ action is known precisely, at
least at long wavelengths, as it may be determined by an anomaly
cancelling mechanism. The gauge part of the WZ action was first determined
for Type I theory in~\cite{CLNY}, while a general D-brane gauge WZ action
was obtained in~\cite{Li,schmid,Douglas}. In particular in~\cite{Douglas}
a compact way of writing the action was given. The first gravitational
term of the WZ action, obtained in~\cite{ber}, was related using a chain
of dualities to a one-loop term discovered in~\cite{VW}. This term
provided further evidence for the Heterotic/Type IIA duality first
suggested in~\cite{Hull}. Later the whole WZ action was determined by an
anomaly cancelling argument~\cite{GHM} (for a review of the WZ
couplings see~\cite{bachas}). While D-branes themselves are non-chiral, a
configuration of intersecting D-branes which contained in its worldvolume
theory chiral fermions was found~\cite{GHM}. Using the anomaly inflow
mechanism~\cite{CH} the relevant anomaly cancelling terms in the action
were obtained. This was generalised to include gravitational couplings to
a non-trivial normal bundle in~\cite{CY}
and~\cite{MM}. The WZ D-brane coupling can be summarised as

\be
I^D_{WZ}=T_p\int_{{\cal B}_p}C\w\tr\exp{\left(iF/(2\pi)\right)}\w\sqrt{{\hat
A}(R)},\label{theone}
\ee

\ni where ${\cal B}_p$ is the world-volume of a Dirichlet $p$-brane,
$T_p^2=2\pi(4\pi^2\alpha^\prime)^{3-p}$ as in~\cite{PolRR}\footnote{In the notation of~\cite{PolRR} we set $\alpha_p=1$.} and
 $F$ is the world-volume gauge field (more generally when the \NSNS
potential $B$ is non-zero $F$ is replaced by ${\cal F}=F-B$).
${\hat A}$ is the Dirac or A-roof genus, and $R$ is the pull-back of the
curvature from the space-time to the D-brane world-volume. $C$ is a sum
over the \RR potentials as differential forms. The worldvolume integral
will pick out terms which appear as $p+1$ forms. The gauge part of this
action was computed by several authors~\cite{ Douglas,Li, schmid} and
can be obtained by constructing an appropriate boundary state and
requiring it to be BRST invariant. In this paper the gravitational part of
the WZ term is discussed, so $F$ is set to zero.  Thus for example the
couplings of the \RR two form are

\be
I_{\mbox{D-brane}}=T_1\int_{{\cal B}_1}\!\!\C2+
\frac{T_5}{24(4\pi)^2}\int_{{\cal B}_5}\!\!\C2\!\!\w\tr(R^2)
+T_9\int_{{\cal B}_9}\!\!
\C2\!\!\w\!\left[\frac{1}{1152(4\pi)^4}(\tr R^2)^2+
\frac{1}{720(4\pi)^4}\tr(R^4)\right].\label{Dbrane}
\ee

Compactifications of Type~II theories involving gauging
worldsheet orientation produce orientifold planes
(O-planes)~\cite{t1orb, DLP, Gimon}. For a review of some such
compactifications see~\cite{Mukhor}. Worldvolume actions of D-branes and
O-planes exhibit some similarities. Both objects carry \RR charge
and have WZ gravitational couplings. An important difference is
that D-branes have a worldvolume gauge field not present on O-planes.  The
gravitational couplings of O-planes were obtained by considering Type~I
theory as an orientifold of Type~IIB~\cite{t1orb}.\footnote{The ten
dimensional case is somewhat degenerate as branes and planes fill
spacetime.} While this paper was in its final stages of
preparation~\cite{MSS} appeared which has substantial overlap
with the
content of this paper. The form of the orientifold WZ coupling is given
by

\be
I^O_{WZ}=-2^{p-4}T_p\int_{{\cal B}_p}C\w\sqrt{L(R/4)},\label{hirz}
\ee

\ni where $L$ is the Hirzebruch polynomial. In~\cite{MSS} the normal
bundle couplings were also obtained. Expanding the above action for the
O9-plane gives

\be
I_{\mbox{O9-plane}}=T_9\int_{{\cal B}_9}-32 C^{(10)}+
\C6\!\!\w\frac{2}{3(4\pi)^2}\tr(R^2)
+\C2\!\!\w\left[\frac{-1}{144(4\pi)^4}(\tr R^2)^2
+\frac{7}{180(4\pi)^4}\tr(R^4)
\right].\label{oplane}
\ee

\ni The first term in the above is simply the 10 dimensional
tadpole~\cite{PC,CLNY-1}. The eight form coupling is different to the one
obtained in~\cite{dasgupta}, but agrees with~\cite{MSS}. The four form
coupling is as postulated by~\cite{dasgupta} which is particularly
reassuring since it is needed for consistency with M-theory~\cite{senm}. 

Apart from~\cite{ber}, very little evidence has been presented in the
literature for the above terms (in particular for the eight form
gravitational couplings). So it is interesting to check their presence
explicitly. We confirm both sets of terms given in the equations above by
calculating the relevant amplitudes in superstring theory. Other \RR
potentials will have similar couplings to the curvature two forms, however
these follow from the above couplings by T-duality.

A recent paper made some initial steps in calculating the gravitational
couplings of D-branes and O-planes~\cite{Craps} by calculating the four
form term of the Dirac genus. In this paper we shall verify the complete
structure of both the D-brane and O-plane actions by considering
appropriate scattering amplitudes in the world-volume of D-branes and
O-planes. More precisely we shall scatter $q$ gravitons and $C^{(p+1)}$ in
the worldvolume of a D$(p+2q)$-brane. This paper is organised as follows.
In Section~\ref{sec2} we discuss the calculations that are to be performed
as well as the conventions used. In sections~\ref{sec3} and~\ref{sec4} the
four and eight form amplitudes for D-branes and crosscaps are calculated. 
The D-brane and O-plane amplitude results are compared with the actions
given above in sections~\ref{sec5} and~\ref{sec6},
respectively. In section~\ref{sec6} we discuss also the Green-Schwarz
anomaly canceling mechanism~\cite{GS}. 

\section{Preliminaries}\label{sec2}

We intend to confirm the gravitational couplings ~(\ref{Dbrane}) 
and~(\ref{oplane}) by calculating scattering amplitudes of the \RR
two-form with two or four gravitons on either a D-brane or O-plane. This
will be done by considering the worldvolume scattering of an appropriate
number of
gravitons with one \RR potential in the presence of a boundary or
crosscap. The boundary state corresponds to a D-brane while the crosscap
to an orientifold plane.  The
boundary state formalism was developed in~\cite{CLNY-1,PC,CLNY} for
purely Neumann boundary conditions. Dirichlet boundary conditions were
discussed in~\cite{Green76}. Boundary states with Dirichlet boundary
conditions were initially constructed for D-instantons in~\cite{Green94}.
Mixed boundary states were first discussed by Polchinski~\cite{PolRR}, and
in~\cite{Li,cK,Douglas,schmid} and have been used extensively since, see
for example~\cite{Billo,diVecchia,senbdr}.

We shall be using vertex operators for gravitons in the $(-1,0)$
and $(0,0)$ pictures, and the \RR two form in the 
$(-\frac{3}{2},-\half)$ picture. These are

\ba
V^g_{(-1,0)}(k,\zeta,z,\bar{z})&=&
\zeta_{\mu\nu}\psi^\mu(z)
(\bar{\partial}\tilde{X}^\nu+\frac{k}{2}\tilde{\psi}\tilde{\psi}^\nu)
(\bar{z})e^{ik\cdot X(z,\bar{z})}, \\
V^g_{(0,0)}(k,\zeta,z,\bar{z})&=& \zeta_{\mu\nu}
(\partial X^\mu+\frac{k}{2}\psi\psi^\mu)(z)
(\bar{\partial}\tilde{X}^\nu+\frac{k}{2}\tilde{\psi}\tilde{\psi}^\nu)
(\bar{z})e^{ik\cdot X(z,\bar{z})}, \\
V^{\C2}_{\left(-3/2,-1/2\right)}(k,\C2,z,\bar{z})&=& \frac{1}{n!}\C2_{\mu\nu} 
S^A\tilde{S}^B\Gamma^{\mu\nu}_{AB} e^{ik\cdot
X(z,\bar{z})}, 
\ea

\ni where ghost and superghost dependence has been suppressed.

The mode expansions for the fields on a cylindrical world-sheet are

\ba
X^\mu(\tau,\sigma)&=&x^\mu+2\pi p^\mu\tau + i\sum_{n\neq 0}\frac{1}{n}
(\alpha^\mu_n e^{-in(\tau-\sigma)}+ \tilde{\alpha}^\mu_n
e^{-in(\tau+\sigma)}),\nonumber \\
\psi^\mu(\tau,\sigma)&=&\sum_r\psi^\mu_r
e^{-in(\tau-\sigma)}, \spc\spc
\tilde{\psi}^\mu(\tau,\sigma)=\sum_r\psi^\mu_r e^{-in(\tau+\sigma)},
\ea

\ni where $0\le\sigma\le2\pi$. The boundary state is taken to be the
end-state at $\tau=0$ of a semi-infinite cylinder, and satisfies

\ba
\left(\partial_\tau X^\mu(\tau=0,\sigma)\right)\bdr &=& 0 \;\;\;\; 
\mu=0,\dots,p,\nonumber \\
X^\mu(\tau=0,\sigma)\bdr &=& 0 \;\;\;\; \mu=p+1,\dots,9,\nonumber \\
(\psi^\mu\pm i\tilde{\psi}^\mu)\bdr &=& 0 \;\;\;\; \mu=0,
\dots,p,\nonumber \\
(\psi^\mu\mp i\tilde{\psi}^\mu)\bdr &=& 0 \;\;\;\;
\mu=p+1,\dots,9,
\ea

\ni or in terms of modes,

\ba
-\alpha^\mu_n\bdr&=&\tilde{\alpha}^\mu_{-n}\bdr, \;\;\;\;
\mu=0,\dots,p,\nonumber \\
\alpha^\mu_n\bdr&=&\tilde{\alpha}^\mu_{-n}\bdr,\;\;\;\;
\mu=p+1,\dots,9,\nonumber \\
\psi^\mu_r\bdr&=&\mp i\tilde{\psi}^\mu_{-r}\bdr, \;\;\;\;
\mu=0,\dots,p,\nonumber \\
\psi^\mu_r\bdr&=&\pm i\tilde{\psi}^\mu_{-r}\bdr, \;\;\;\; \mu=p+1,\dots,9.
\label{eq10}
\ea

\ni In the above $r$ is half integer in the \NSNS sector and integer in
the \RR sector. 

The relevant boundary conditions for a crosscap are~\cite{CLNY-1,PC}

\ba
\frac{\partial}{\partial\tau}X^\mu(\sigma+\pi,\tau=0)
&=&-\frac{\partial}{\partial\tau}X^\mu(\sigma,\tau=0)\nonumber \\
X^\mu(\sigma+\pi,\tau=0)&=&X^\mu(\sigma,\tau=0).\label{cc1}
\ea

\ni In constructing O$p$-planes one compactifies $9-p$ directions on
circles and performs T-duality in these directions~\cite{t1orb, DLP}.  The
boundary conditions in the compact directions then read

\be
X^\mu(\sigma+\pi,\tau=0)=-X^\mu(\sigma,\tau=0).\label{cc2}
\ee

\ni This fixes the transverse position of the Orientifold plane at
$x^\mu=0$. The effect of a crosscap state on modes is given by
equation~(\ref{eq10}) but with an extra $(-1)^n$ factor inserted. 

\ni The boundary and crosscap state are constructed as eigenvectors and as
such do not have a fixed normalisation. However, one may normalise the
crosscap relative to the boundary state by looking at the factorisation of
cylinder and M\"obius strip diagrams in the open string
channel~\cite{PC,CLNY-1}. We denote by $\eta_{10}$ the normalisation of a crosscap with all Neumann boundary conditions relative to the boundary state of a D9-brane.

We shall adopt here the fermionic zero
mode conventions of~\cite{Billo} for the D$p$-brane boundary state,

\be
\left|B_\psi\right>^{(0)}={\cal
M}_{AB}\left|A\right>\left|\right.\!\!{\tilde
B}\!\!\left.\right>,
\ee

\ni where

\be
{\cal M}=\frac{T_p}{32}
(C\Gamma^0\dots\Gamma^p)\left(1\pm i\Gamma^{11}\right).
\ee

\ni Here $C$ is the charge conjugation matrix and
$A,B,\dots$ are 32-dimensional indices for spinors in 10 dimensions. The
fermionic zero modes are represented by the following action

\be
d_0^\mu\left|A\right>\left|\right.\!\!{\tilde
B}\!\!\left.\right>=
\frac{1}{\sqrt{2}}(\Gamma^\mu)^A_{\,\,C}(1)^B_{\,\,D}\left|C\right>
\left|\right.\!\!{\tilde D}\!\!\left.\right>\label{ferzer}
\ee

\ni The right moving fermions' representation can be obtained from the
boundary state's effect on the fermionic modes. 

The $C^{(p)}$ state with momentum $k$ is placed at $\tau=\infty$ and is
given by

\be
\left<\right.\!\!C^{(p)},k\left|\right.=\frac{1}{p!}
C_{\mu_1\dots\mu_{p}}\Gamma^{\mu_1\dots\mu_{p}}_{A{\tilde 
B}}\left<\right.\!\!A\left|\right.\otimes\left<\right.\!\!{\tilde
B}\left|\right.\otimes\left<\right.\!\!k\left|\right. .\label{rrstate}
\ee

\ni With the above conventions

\be
\left<\right.\!\!C^{(p)},k\bdr=\frac{T_p}{p!}\varepsilon^{\mu_1\dots\mu_{p+1}}
C^{(p)}_{\mu_1\dots\mu_{p+1}},
\ee

\ni in agreement with~(\ref{theone}).

\section{The four form amplitudes}\label{sec3}

In this section we discuss the $\tr R^2$ WZ term in the action. The first
part shows in some detail how the D-brane calculation is performed while
the second part of the section obtains the O-plane amplitude in a
straightforward generalisation of the D-brane calculation. 

\subsection{The D5-brane amplitude}

We are interested in scattering two gravitons and a \RR two form, $\C2$,
in the presence of a D5-brane in order to verify the D-brane worldvolume
term

\be
\frac{T_5}{24(4\pi)^2}\int_{{\cal B}_5}\C2\tr R\w R,
\ee

\ni in the low energy effective action. We do this by choosing a
worldsheet time coordinate $\tau$ and a periodic space coordinate
$\sigma$. Furthermore, we fix the position of the $\C2$ vertex operator
and that of the D5-brane. In effect we need to compute

\ba
A_2&=&\ck V_g(k^1)\Delta V_g(k^2)\Delta\bdr\nonumber \\
&=&(4\pi)^{-2}\int_{\modz\ge 1}\frac{d^2z}{\modz^2}\int_{\mody\ge 1}
\frac{d^2y}{\mody^2}
\ck T\left\{V_g\left(k^1,y,\bar{y}\right)
V_g\left(k^2,z,\bar{z}\right)\right\}\bdr,\label{2gr}
\ea

\ni where $V_g$ is a graviton vertex operator in an appropriate picture,
$\Delta$ is a closed string propagator and $T\{\dots\}$ indicates time
ordering. In this calculation all momenta shall be restricted to the
worldvolume directions, and we denote by $M_2$ the matrix element above.

On the disc the superghost number anomaly is -2.  Thus we can work with
the gravitons in the $(0,0)$ picture and the \RR two form in the
$\left(-\half,-\frac{3}{2}\right)$ picture. Alternately, the \RR two form
can be in the $\left(-\half,-\half\right)$ picture with one of the
gravitons in the $(-1,0)$ picture while the other in the $(0,0)$
picture.

Turning to the ghost factor, BRST invariance of the boundary state fixes
the ghost zero mode piece of $\bdr$ to be
$\half\left(c_0+\tilde{c}_0\right)\left|\downarrow\downarrow\right>$.  The
propagator next to the boundary has to have a ghost piece
$\half(b_0+\tilde{b}_0)$ since the cylinder has only one Teichm\"uller
parameter. The other propagator will contain the usual $\oint
b(z)\oint\tilde{b}(\bar{z})$ factor while the vertices will have factors
of $c\tilde{c}$ in order to make them BRST invariant. Thus the state $\ck$
will have a $c_{-1}\tilde{c}_{-1}$ ghost piece. Combining these we see
that the ghost factor is

\be
\left<\Omega\right|c_{-1}\tilde{c}_{-1}c(z)\tilde{c}(\bar{z})
\left|\downarrow\downarrow\right> 
= \left<\Omega\right|c_{-1}\tilde{c}_{-1}c(z)\tilde{c}(\bar{z})
c_{1}\tilde{c}_{1}\left|\Omega\right> = 1.
\ee

\ni At this point we note that the ghost piece will be the same for the 4
graviton scattering discussed in the next section. In fact the ghost piece
for any tree level process in the cylinder channel will simply be equal
to $1$. 

The fermionic zero modes play an important role in the calculation. The
boundary state contribution amounts to a factor of
$\Gamma^0\dots\Gamma^5$. Furthermore, the right-moving spin field
$\tilde{S}^\alpha$ gets converted into a left-moving spin field by the
boundary state, resulting in a trace over the Clifford algebra. We will
have to saturate the $\Gamma^0\dots\Gamma^5$ factor by picking out terms
with precisely 6 fermionic zero modes in them. Since we are dealing with
massless, physical states we have\footnote{A symmetric traceless
polarisation tensor $\zeta_{\mu\nu}$ can be expressed as
$\zeta_{\mu\nu}=\sum_{i=1}^n a_i\zeta^i_\mu\zeta^i_\nu$ with $a_i$
constant. Without loss of generality we take
$\zeta_{\mu\nu}=\zeta_\mu\zeta_\nu$.} $k^2=k\cdot\zeta=0$, we can only
select precisely two zero modes from each graviton vertex, one contracted
into $\zeta$ and the other into $k$.\footnote{This follows from the fact
that $\zeta\!\!\cdot\!\Gamma \zeta\!\!\cdot\!\Gamma=k\!\cdot\!\Gamma
k\!\cdot\!\Gamma=0$.} The $\C2$ state contributes a further two gamma
matrices. Thus the fermion zero modes contribute a factor

\be
W=\frac{32}{4}\varepsilon^{\mu_1\dots\mu_6}
\C2_{\mu_1\mu_2}\zeta^1_{\mu_3}k^1_{\mu_4}\zeta^2_{\mu_5}k^2_{\mu_6},
\ee

\ni where $32$ and $4$ come from the gamma matrix trace and from the
representation of the fermionic zero modes as in equation~(\ref{ferzer}),
respectively. 

As a result of the above discussion we obtain modified graviton vertices
in the $(0,0)$ and $(-1,0)$ pictures, which subject to the above zero mode
factor are

\ba
V^M_{(-1,0)}(k,\zeta,z,\bar{z})&=&
-\frac{\zeta_\mu}{2}(\psi^\mu\pm
i\tilde{\psi}^\mu)e^{ikX}=-\frac{\zeta_\mu}{2}\Psi^\mu e^{ikX},
\nonumber \\
V^M_{(0,0)}(k,\zeta,z,\bar{z})&=&
\left[-\zeta_\mu(\partial X^\mu+\frac{k_\nu}{2}\psi^\nu\psi^\mu)+
\zeta_\mu({\bar\partial}\tilde{X}^\mu+\frac{k_\nu}{2}\tilde{\psi}^\nu
\tilde{\psi}^\mu)+\frac{\pm i}{2}(\zeta_\mu\psi^\mu
k_\nu\tilde{\psi}^\nu-k_\mu\psi^\mu\zeta_\nu\tilde{\psi}^\nu)
\right]e^{ikX}\nonumber \\
&=&-\zeta_\mu(\partial^*{\cal
X}^\mu+\frac{k_\nu}{2}\Psi^\nu\Psi^\mu)e^{ikX}=
P(k,\zeta,z,{\bar z})e^{ikX}.
\label{modver}
\ea

\ni where the prefactor $P(k,\zeta,z,{\bar z})$ denotes the term in square
brackets in the above equation. Here $\Psi^\mu(z,{\bar z})=\psi^\mu(z)\pm
i{\tilde\psi}^\mu({\bar z}),\;\; {\cal X}^\mu(z,{\bar z})=X^\mu(z)+{\tilde
X}^\mu({\bar z})$ and $\partial^*=\partial-{\bar\partial}$. The fact that
we can split the amplitude in this way makes the computation substantially
easier as we are now dealing with half as many multiplicative factors. We
define

\ba
a&=&\frac{z}{y-z}, \spc \bar{a}=\frac{\bar{z}}{\bar{y}-\bar{z}},
\spc A=\bar{a}-a=\frac{\mbox{Im}(z\bar{y})}{\left|z-y\right|^2},
\nonumber \\
b&=&\frac{1}{\bar{y}z-1}, \spc \bar{b}=\frac{1}{y\bar{z}-1},
\spc B=\bar{b}-b=\frac{\mbox{Im}(z\bar{y})}{\left|z\bar{y}-1\right|^2}.
\ea

\ni The matrix element in equation~(\ref{2gr}) can then be written as

\be
M_2=\frac{T_5W}{4}\ck P_1\e1 P_2\e2\bdr=
\frac{T_5W}{4}\ck(P_1+A_1)(P_2+A_2)\bdr{\cal E},
\ee

\ni where $P_1=P(k^1,\zeta^1,y,{\bar y}),\spc P_2=P(k^2,\zeta^2,z,{\bar
z})$ and

\be
A_1=\frac{\zeta^1\cdot k^2}{2}(A-B),\spc
A_2=-\frac{\zeta^2\cdot k^1}{2}(A-B),\spc
{\cal E}=e^{\left[ik^1\cdot X^-,ik^2\cdot X^*\right]}.
\ee

\ni $X^-$ is the annihilation part of $X$ while $X^*$ is $X$ with all
annihilation operators converted to creation operators using the boundary 
state.

The prefactor $P_2$ may now be bounced off the boundary state converting
it into $P_2^*$ which contains only creation operators. The relevant
prefactor commutator is 

\be
\left[P_1,P_2^*\right]
=H_{12}+G_{12},
\ee

\ni where $G_{12}$ and $H_{12}$ are given by

\ba
G_{12}&=&
-\frac{1}{4}(A+B)k^1_{[\sigma}\zeta^1_{\mu]}k^{2[\mu}\zeta^2_{\rho]}
\Psi^\rho\Psi^\sigma,\\
H_{12}&=&
\zeta^1\cdot\zeta^2\left(\frac{zy}{(z-y)^2}+
\frac{z\bar{y}}{(z\bar{y}-1)^2}+c.c.\right)+\frac{1}{8}
\zeta^1_{[\mu}k^1_{\nu]}\zeta^{2[\nu}k^{2\mu]}
(A+B)^2,
\ea

\ni where $\zeta^1_{[\mu}k^1_{\nu]}=\zeta^1_\mu k^1_\nu-\zeta^1_\nu
k^1_\mu$ . Since $G_{12}$ is normal ordered it will not
contribute to the two graviton scattering amplitude.  ${\cal E}$ is given
by

\be
\left(y-z\right)^{k^1\cdot k^2/4}\left(y-\frac{1}{\bar{z}}\right)^{k^1\cdot k^2/4}y^{-k^1\cdot k^2/4}\times c.c.
\ee

\ni The total matrix element is

\be
M_2=\frac{T_5W}{8}\left\{H_{12}-\frac{1}{4}\zeta^1\cdot k^2\zeta^2\cdot 
k^1(A-B)^2\right\}|y-z|^{k^1\cdot k^2/2}
|y-\frac{1}{\bar{z}}|^{k^1\cdot k^2/2}|y|^{-k^1\cdot k^2/2}.
\ee

The amplitude as it stands contains second order poles and thus would
appear to factorise on tachyonic excitations. These undesirable poles are
expected when using vertex operators in certain
pictures~\cite{GSei}.\footnote{For a recent addition to this reasoning in
the context of D-branes see~\cite{Gutpole}.} It can be shown that they
combine into terms which contain first order poles and total derivatives.
Performing this operation the amplitude under consideration becomes

\be
A_2=\frac{T_5W}{128\pi^2}
\int_{\modz\ge 1} \frac{d^2z}{\modz^{2+k^1\cdot k^2/2}}\int_{\mody\ge 1} 
\frac{d^2y}{\mody^{2+k^1\cdot k^2/2}}
\frac{(k^1\cdot\zeta^2 k^2\cdot\zeta^1-k^1\cdot k^2\zeta^1\cdot\zeta^2)
\mbox{Im}^2(z\bar{y})}{|z-y|^{2-k^1\cdot k^2/2}|z\bar{y}-1|^{2-k^1\cdot k^2/2}}.
\ee

\ni Since we are concerned with scattering in the world-volume of D-branes
the transverse momenta vanish. In that case momentum conservation implies
that $k^i\cdot k^j=0$ and so ${\cal E}=1$. The region of integration can
be divided into two sub-regions - $\mody<\modz$ and $\modz<\mody$. In each
of these regions one may expand $a,b$ and their complex conjugates in a
power series. This gives

\ba
& &\int_{\modz\ge 1} \frac{d^2z}{\modz^2}\int_{\mody\ge 1} \frac{d^2y}{\mody^2}AB
=2\int_{\modz\ge 1} \frac{d^2z}{\modz^2}\int_{\modz\ge\mody\ge 1} \frac{d^2y}{\mody^2}AB
\nonumber \\
&=&2\int_{r= 1}^{\infty} \frac{drd\theta_1}{r}\int_{\rho= 1}^{\infty} 
\frac{d\rho d\theta_2}{\rho}
\sum_{m,n=1}^{\infty}\left(\frac{\rho}{r}\right)^m
(e^{i(\theta_1-\theta_2)m}-e^{i(\theta_2-\theta_1)m})
\left(\frac{1}{r\rho}\right)^n
(e^{i(\theta_2-\theta_1)n}-e^{i(\theta_1-\theta_2)n}),\label{int4}
\ea

\ni where $y=\rho e^{i\theta_2}$ and $z=re^{i\theta_1}$. Integrating out
the angular variables one obtains

\be
4(2\pi)^2\int_{r=1}^{\infty} \int_{\rho=1}^{r} 
\sum_{n=1}^{\infty}r^{-2n}
\frac{d\rho}{\rho}\frac{dr}{r}
=4(2\pi)^2\sum_{n=1}^{\infty}\int_{r=1}^{\infty} r^{-2n-1}\ln r dr.
\label{int1}
\ee

\ni Since

\be
\int_{r=1}^{\infty}r^{-2n-1}\ln rdr=\frac{1}{4n^2},
\ee

\ni the integral in equation~(\ref{int4}) becomes

\be
4\pi^2\sum_{n=1}^{\infty}\frac{1}{n^2}=
4\pi^4\left|B_2\right|=\frac{2\pi^4}{3},\label{int2}
\ee

\ni where $B_n$ is the $n$-th Bernoulli number. We may then write down the
final amplitude

\be
A_2=\frac{T_5\pi^2}{192}\varepsilon^{\mu_1\dots\mu_6}
\C2_{\mu_1\mu_2}k^1_{\mu_3}\zeta^1_{\mu_4}k^2_{\mu_5}\zeta^2_{\mu_6}
(\zeta^1\cdot k^2\zeta^2\cdot k^1-k^1\cdot k^2\zeta^1\cdot\zeta^2),\label{4dim}
\ee

\ni We observe that the evaluation of the integral differs by a factor of
2 from the one in the work of Craps and Roose~\cite{Craps}.\footnote{This
may account for the discrepancy that these authors reported between the
amplitude and the low energy effective action.}

\subsection{The O9-plane amplitude}\label{sec32}

In this subsection we consider scattering $\C6$ with two gravitons on a
O9-plane. The crosscap state which represents the O9-plane has a zero mode
contribution of the form $\Gamma^0\dots\Gamma^9$, while the $\C6$ state
also has six gamma matrices. The fermionic zero mode contribution to the
crosscap amplitude is then

\be
W=\frac{32}{4}\varepsilon^{\mu_1\dots\mu_{10}}\C6_{\mu_1\dots\mu_6}
k^1_{\mu_7}\zeta^1_{\mu_8}k^2_{\mu_9}\zeta^2_{\mu_{10}}.
\ee

\ni The crosscap state is much like the boundary state except that it
contains an extra $(-1)^n$ in the exponential terms of the coherent states
describing the boundary state.  This modifies $B$ to

\be
B^\prime=\frac{-1}{1+y\bar{z}}-\frac{-1}{1+z\bar{y}},
\ee

\ni while $A$ remains unchanged. Thus the final integration now
includes a factor of $(-1)^n$. It is easy to sum the appropriate series
using

\be
\sum_{k=1}^{\infty}(-1)^{k}\frac{1}{k^{2n}}=\frac{(1-2^{2n-1})
\pi^{2n}}{(2n)!}|B_{2n}|.\label{sums}
\ee

\ni The amplitude for scattering a $\C6$ and two gravitons off a boundary
can be obtained as a generalisation of the calculation in the previous
section. It follows from equation~(\ref{sums}) that the crosscap amplitude
is $-\half\eta_{10}$ times the boundary amplitude. 

\section{The eight-form amplitudes}\label{sec4}

In this section the eight-form gravitational couplings of D-branes and
O-plane states are studied.  We first discuss the D9-brane amplitude and
then, as before, by a simple modification obtain the O9-plane amplitude.
This will allow us to explicitly check gravitational anomaly cancellation
in the ten dimensional Type I theory. 

\subsection{The D9-brane amplitude}

In order to confirm the ten-dimensional term we calculate the amplitude
for four gravitons and one $\C2$ on a D9-brane. This is given by

\be
A_4=(4\pi)^{-4}\int_{\mdz4\ge 1} \frac{d^2z_4}{\mdz4^2}\int_{\mdz3\ge
1}\frac{d^2z_3}{\mdz3^2}
\int_{\mdz2\ge 1}\frac{d^2z_2}{\mdz2^2}\int_{\mdz1\ge
1}\frac{d^2z_1}{\mdz1^2}M_4,\label{mesf}
\ee

\ni where 

\be
M_4=\ck
T\left\{V_g\left(k^1,z_1,\bar{z}_1\right)V_g\left(k^2,z_2,\bar{z}_2\right)
V_g\left(k^3,z_3,\bar{z}_3\right)V_g\left(k^4,z_4,\bar{z}_4\right)\right\}\bdr.
\ee

\ni As in the D5-brane case the \RR
two form induces a trace over the fermionic zero modes of the boundary
state and the vertex operators. This contributes a factor

\be
W=\frac{32}{16}\varepsilon^{\mu_1\dots\mu_{10}}\C2_{\mu_1\mu_2}
\zeta^1_{\mu_3}k^1_{\mu_4}\dots\zeta^4_{\mu_9}k^4_{\mu_{10}},
\ee

\ni and allows for the definition of modified graviton vertex operators as
before. Defining

\ba
a_{ij}&=&\frac{z_j}{z_i-z_j},\spc b_{ij}=\frac{1}{\bar{z}_iz_j-1},
\nonumber \\
A_{ij}&=&\frac{\mbox{Im}(z_i\bar{z}_j)}{\left|z_i-z_j\right|^2},\spc
B_{ij}=\frac{\mbox{Im}(z_i\bar{z}_j)}{\left|z_i\bar{z}_j-1\right|^2}.
\ea

\ni the matrix element may be written as

\ba
M_4&=&\frac{T_9W}{16}\ck P_1\e1 P_2\e2 P_3\e3 P_4\e4\bdr \nonumber \\
&=&\frac{T_9W}{64}{\cal E}
\ck(P_1+A_1)(P_2+A_2)(P_3+A_3)(P_4+A_4)\bdr,\label{fourma}
\ea

\ni where the $P_i$ are defined as previously and  

\be
A_i=\sum_{j\neq i}\frac{\zeta^i\cdot k^j}{2}(A_{ij}-B_{ij}).
\ee

\ni ${\cal E}$ is given by

\be {\cal E}=\prod_{\begin{array}{l}\\[-15.5pt]\mbox{{\scriptsize$
i=1$}} \\[-4pt]
\mbox{{\scriptsize $j>i$}}\end{array}}^{4}(z_i-z_j)^{k^i\cdot k^j/4}
(z_i-\frac{1}{\bar{z}_j})^{k^i\cdot k^j/4}z_i^{-k^i\cdot k^j/4}\times c.c.
\ee

\ni The computation of the matrix element uses the commutator
$\left[P_i,P_j^*\right]$ obtained previously, where now the $G_{ij}$
operator has a non-trivial effect. The overall matrix element is

\ba
\!\!\!\!\!\!\!\!\!\!\!M_4&\!\!\!\!=&
\!\!\!\!
\frac{T_9W}{256}\!\!\left[\tr(R^1R^2)\tr(R^3R^4)A_{12}B_{12}A_{34}B_{34}+
\tr(R^1R^3)\tr(R^2R^4)A_{13}B_{13}A_{24}B_{24}+
\tr(R^2R^3)\tr(R^1R^4)A_{23}B_{23}A_{14}B_{14}\right]{\cal E}
\nonumber \\
& &
\!\!\!\!+\frac{T_9W}{512}\tr(R^1R^2R^3R^4)\left[(A_{14}B_{12}+
A_{12}B_{14})(A_{23}A_{34}+B_{34}B_{23})
+(A_{23}B_{34}+A_{34}B_{23})(A_{12}A_{14}+B_{12}B_{14})\right]{\cal E}
\nonumber \\
& &
\!\!\!\!+\frac{T_9W}{512}\tr(R^1R^3R^2R^4)\left[(A_{14}B_{13}+
A_{13}B_{14})(A_{23}A_{24}+B_{24}B_{23})+
(A_{23}B_{24}+A_{24}B_{23})
(A_{13}A_{14}+B_{13}B_{14})\right]{\cal E}
\nonumber \\
& &
\!\!\!\!+\frac{T_9W}{512}\tr(R^1R^3R^4R^2)\left[(A_{12}B_{13}+
A_{13}B_{12})(A_{34}A_{24}+B_{24}B_{34})
+(A_{34}B_{24}+A_{24}B_{34})(A_{13}A_{12}+B_{13}B_{12})\right]
{\cal E},\label{befint}
\ea

\ni which we integrate over the measure in equation~(\ref{mesf}). Here the
traces are expressed in terms of the momenta and polarisations of the
gravitons (see equation~(\ref{grpol}) below).

The integral of the first term (the $\tr(R^2)^2$ term) in~(\ref{befint}) 
splits into a product of two integrals encountered in the previous
section. The integration gives a factor of $(2\pi^4/3)^2=4\pi^8/9$. Now
consider the $\tr(R^1R^2R^3R^4)$ term. This contains two kinds of
integrals:  ones involving three $A$'s and one $B$ and ones with three
$B$'s and one $A$. It is easy to see by manipulating the indices on the
$z_i$, that each of the four AAAB terms contributes the same amount. 
Similarly all the BBBA integrals are the same. Let us then integrate
first, say, $A_{12}B_{14}B_{23}B_{34}$. As in the previous section we
split the region of integration into two parts namely
$\left|z_1\right|>\left|z_2\right|$ and
$\left|z_1\right|<\left|z_2\right|$. We then expand in a series of
$z_2/z_1$ or $z_1/z_2$. After integrating out the angular coordinates
$\theta_i$, where $z_i=x_ie^{i\theta_i}$, the two regions again combine to
give

\ba
& &\prod_{i=1}^{4}\int_{|z_i|\ge 1}\frac{d^2z_i}{|z_i|^2}
A_{12}B_{14}B_{23}B_{34}
=64\pi^4\!\!\int_{x_2,x_3,x_4=1}^{\infty}\int_{x_1=1}^{x_2}
\sum_{n=1}^{\infty}\left(x_2x_3x_4\right)^{-2n}\frac{dx_4dx_3dx_1dx_2}
{x_4x_3x_1x_2}
\nonumber \\& &
=16\pi^4\sum_{n=1}^{\infty}
\frac{1}{n^2}\int_{x_2=1}^{\infty}\!\!\!\!x_2^{-2n-1}\ln x_2dx_2
=4\pi^4\sum_{n=1}^{\infty}\frac{1}{n^4}=\frac{2\pi^8}{45}.\label{int3}
\ea

\ni The AAAB integrals are a bit more involved as now one has 8 regions,
but again they can be integrated fairly easily to give for each AAAB
integral $14\pi^8/45$. The final amplitude then becomes

\ba
A_4&\!\!\!\!=&\!\!\!\!
\frac{T_9\pi^4}{73728}\varepsilon^{\mu_1\dots\mu_{10}}
\C2_{\mu_1\mu_2}k^1_{\mu_3}\zeta^1_{\mu_4}\dots
k^4_{\mu_9}\zeta^4_{\mu_{10}}\zeta^1_{[\mu_1}k^1_{\mu_2]}
\nonumber \\& &\!\!
\times\left[
\zeta^{2[\mu_2}k^{2\mu_1]}
\zeta^3_{[\mu_3}k^3_{\mu_4]}\zeta^{4[\mu_4}k^{4\mu_3]}
+\zeta^{3[\mu_2}k^{3\mu_1]}
\zeta^2_{[\mu_3}k^2_{\mu_4]}\zeta^{4[\mu_4}k^{4\mu_3]}
+\zeta^{4[\mu_2}k^{4\mu_1]}
\zeta^2_{[\mu_3}k^2_{\mu_4]}\zeta^{3[\mu_4}k^{3\mu_3]}\right]
\nonumber \\
& &
+\frac{T_9\pi^4}{46080}\varepsilon^{\mu_1\dots\mu_{10}}
\C2_{\mu_1\mu_2}k^1_{\mu_3}\zeta^1_{\mu_4}\dots
k^4_{\mu_9}\zeta^4_{\mu_{10}}\zeta^1_{[\mu_1}k^1_{\mu_2]}\nonumber \\
& &\!\!
\times\left[
\zeta^{2[\mu_2}k^{2\mu_3]}
\zeta^3_{[\mu_3}k^3_{\mu_4]}\zeta^{4[\mu_4}k^{4\mu_1]}
+\zeta^{3[\mu_3}k^{2\mu_3]}
\zeta^2_{[\mu_3}k^2_{\mu_4]}\zeta^{4[\mu_4}k^{4\mu_1]}
+\zeta^{3[\mu_2}k^{3\mu_3]}
\zeta^4_{[\mu_3}k^4_{\mu_4]}\zeta^{2[\mu_4}k^{2\mu_1]}\right].\label{8amp}
\ea

\ni As we will see below the symmetry factors of the two eight form terms
are the same, thus we may compare their ratio

\be
\frac{\tr R^4}{(\tr R^2)^2}=\frac{8}{5},
\ee

\ni which is precisely the one expected for the expansion of $\sqrt{{\hat
A}}$ in equation~(\ref{Dbrane}). 

\subsection{The O9-plane amplitude}\label{sec42}

As in the case of the $\tr R^2$ coupling we now calculate the
amplitude involving the \RR two-form, four gravitons and a crosscap. In
effect we are calculating the contribution of a process on a
non-orientable worldsheet to the Green-Schwarz gravitational anomaly
cancelling term. Again there are only minor changes to the amplitude given
in equation~(\ref{befint}). Namely the $B_{ij}$ become

\be
B^\prime_{ij}=\frac{-1}{1+z_i\bar{z}_j}-\frac{-1}{1+z_j\bar{z}_i},
\ee

\ni while the $A_{ij}$ remain fixed. Equations~(\ref{int1})-(\ref{int2})
and~(\ref{int3}) show that the coefficients of $(\tr R^2)^2$ and $\tr R^4$
in the boundary state calculation are proportional to
$\left(\sum\frac{1}{n^2}\right)^2$ and $\sum\frac{1}{n^4}$, respectively.
The crosscap, as discussed above introduces a factor of $(-1)^n$ into
these sums. Using equation~(\ref{sums}) we have

\ba
\left(\sum_{n=1}^{\infty}\frac{(-1)^n}{n^2}\right)^2&=&
\frac{1}{4}\left(\sum_{n=1}^{\infty}\frac{1}{n^2}\right)^2, \nonumber \\
\sum_{n=1}^{\infty}\frac{(-1)^n}{n^4}&=&
-\frac{7}{8}\sum_{n=1}^{\infty}\frac{1}{n^4}.
\ea

\ni Thus, subject to the crosscap normalisation, the crosscap $(\tr
R^2)^2$ and $\tr R^4$ terms are $\frac{\eta_{10}}{4}$ and
$-\frac{7\eta_{10}}{8}$ times the D-brane couplings, respectively. 
Since the symmetry factors of the two eight form terms
are the same we compare their ratio

\be
\frac{\tr R^4}{(\tr R^2)^2}=-\frac{28}{5},
\ee

\ni which is precisely the one expected for the expansion of
$\sqrt{L(R/4)}$ in equation~(\ref{oplane}). 

\section{Comparison with the D-brane action}\label{sec5}

In order to confirm the WZ D-brane action in equation~(\ref{Dbrane}) it is
necessary to write it in a form comparable with the amplitude calculations
of the previous sections. The linearised Ricci two-form in terms of
graviton polarisation and momentum is given by

\be
R_{\mu\nu}^{\;\;\;ab}=(\zeta_\mu k_\nu -k_\mu\zeta_\nu)
(\zeta^a k^b -k^a\zeta^b),\label{grpol}
\ee

\ni where $a,b$ are tangent bundle indices.

The symmetry factors of the $\tr R^2$ term is 2 while both the $\tr R^4$
and $(\tr R^2)^2$ terms have a symmetry factor of 8. The D-brane WZ action
then gives the following amplitudes

\ba
& &\frac{T_5}{48(4\pi)^2}
\varepsilon^{\mu_1\dots\mu_6}\C2_{\mu_1\mu_2}\zeta^1_{\mu_3} k^1_{\mu_4}\zeta^2_{\mu_5} k^2_{\mu_6}
(\zeta^1\!\!\cdot\! k^2\zeta^2\!\!\cdot\! k^1-\zeta^1\!\!\cdot\!\zeta^2 k^1\!\!\cdot\! k^2)\nonumber \\
& &+\frac{T_9}{4608(4\pi)^4}
\varepsilon^{\mu_1\dots\mu_{10}}\C2_{\mu_1\mu_2}\zeta^1_{\mu_3} k^1_{\mu_4}\dots\zeta^4_{\mu_9} k^4_{\mu_{10}}
\zeta^1_{[\mu_1}k^1_{\mu_2]}\nonumber \\
& &\times\left[\zeta^{2[\mu_2}k^{2\mu_1]}
\zeta^3_{[\mu_3}k^3_{\mu_4]}\zeta^{4[\mu_4}k^{4\mu_3]}
+\zeta^{3[\mu_2}k^{3\mu_1]}
\zeta^2_{[\mu_3}k^2_{\mu_4]}\zeta^{4[\mu_4}k^{4\mu_3]}
+\zeta^{4[\mu_2}k^{4\mu_1]}
\zeta^2_{[\mu_3}k^2_{\mu_4]}\zeta^{3[\mu_4}k^{3\mu_3]}\right]
\nonumber \\
& &+\frac{T_9}{2880(4\pi)^4}\varepsilon^{\mu_1\mu_2\dots\mu_9\mu_{10}}
\C2_{\mu_1\mu_2}\zeta^1_{\mu_3} k^1_{\mu_4}\dots\zeta^4_{\mu_9} k^4_{\mu_{10}}\zeta^1_{[\mu_1}k^1_{\mu_2]}\nonumber \\
& &\times\left[\zeta^{2[\mu_2}k^{2\mu_3]}
\zeta^3_{[\mu_3}k^3_{\mu_4]}\zeta^{4[\mu_4}k^{4\mu_1]}
+\zeta^{3[\mu_2}k^{3\mu_3]}
\zeta^2_{[\mu_3}k^2_{\mu_4]}\zeta^{4[\mu_4}k^{4\mu_1]}
+\zeta^{3[\mu_2}k^{3\mu_3]}
\zeta^4_{[\mu_3}k^4_{\mu_4]}\zeta^{2[\mu_4}k^{2\mu_1]}\right].
\label{expected}
\ea

\ni Let us now compare this with the amplitudes calculated in the previous
sections. While the exact normalisation of graviton vertex operators can
be obtained from first principles, this seems unnecessarily convoluted.
Instead we will show that the terms calculated in the amplitudes above are
in the correct ratios to each other.

The graviton
normalisation (say $v$) is obtained by comparing coefficients in
equations~(\ref{4dim}) and~(\ref{expected}) giving $v=(2.\pi^2)^{-1}$. Now
the $\tr(R^2)^2$ coefficient in the amplitude is

\be
\frac{\pi^4v^4T_9}{73728}=\frac{T_9}{4608(4\pi)^4},
\ee

\ni which corresponds to precisely the result expected in
equation~(\ref{expected}). Similarly one can confirm the $\tr(R^4)$
term's coefficient.

\section{The O-Plane action and Green-Schwarz anomaly cancellation}\label{sec6}

In the course of this calculation it became clear that the gravitational
couplings of orientifold planes were somewhat different from the ones
postulated in~\cite{dasgupta}. 

Type I theory in 10 dimensions has a tadpole cancellation due to its gauge
group being SO(32)~\cite{PC,CLNY-1} giving $\eta_{10}=-32$. In lower
dimensions the number of O-planes doubles, thus $\eta_d=-2^{d-5}$. Since
in sections~\ref{sec3} and~\ref{sec4} we expressed the O-plane couplings
as fractions of the D-brane couplings it is now straightforward to confirm
the couplings in equation~(\ref{oplane}). Explicitly the four form
coupling of the O9-plane was $-\eta_{10}/2=16$ times that of the
D9-brane, while the eight-form $\tr R^4$ and $(\tr R^2)^2$ couplings are
$\eta_{10}/4=-8$ and $-7\eta_{10}/8=-28$, respectively times the D9-brane
coupling. These are in exact agreement with equation~(\ref{oplane}).
Gravitational couplings of other O-planes can be similarly deduced and are
summarised in~(\ref{hirz}).

Finally one can confirm the Green-Schwarz gravitational anomaly cancelling
terms~\cite{GS} of Type I theory. These are

\be
S_{GS}=T_9\int \frac{2}{(4\pi)^2}\C6\!\!\w\!\tr
R^2+\frac{1}{12(4\pi)^4}\C2\!\!\w\!\!
\left[\frac{1}{4}\tr(R^2)^2+\tr R^4\right].
\ee

\ni It is a crucial consistency check that the gravitational couplings of
32 D9-branes and an O9-plane reproduce the Green-Schwarz terms. In fact
again exact agreement is found.

\section*{Acknowledgements}

I would like to thank Michael Green for introducing me to this problem and
for his continued support throughout its completion. I would also like to
acknowledge the many useful discussions with Matthias Gaberdiel and Marco
Barrozo. 

\section*{Added note concerning O-planes and the Hirzebruch polynomial}

Following the interpretation of~(\ref{oplane}) in terms of the square root
of the Hirzebruch polynomial~\cite{MSS} it is very plausable that the
O-plane WZ action can be determined from a generalisation of the anomaly
inflow argument that was used in~\cite{GHM} to determine the D-brane WZ
term. For example, two O7-planes intersect in a manner that gives rise to
a chiral 5+1 dimensional $N=1$ supersymmetric theory in the intersection
domain.  The spectrum includes a second rank antisymmetric self-dual
tensor which has an anomaly obtained by descent from the eight form piece
of the Hirzebruch polynomial.

\pagebreak


\begin{thebibliography}{99}

\bibitem{PolRR} J. Polchinski, {\em Dirichlet-Branes and Ramond-Ramond
Charges}, \prl75(1995)4724,  \xxx9510017;

\bibitem{Leigh} R.G. Leigh, {\em Dirac-Born-Infeld Action from Dirichlet
Sigma Model}, \mpla4(1989)2767;

\bibitem{CLNY} C.G. Callan, jr., C. Lovelace, C.R. Nappi and S.A. Yost,
{\em Loop Corrections to Superstring Equations of Motion},
\npb308(1988)221;

\bibitem{Li} M. Li, {\em Boundary States of D-branes and Dy-Strings},
\npb460(1996)351, \xxx9510161;

\bibitem{schmid} C. Schmidhuber, {\em D-Brane Actions}, \npb467(1996)146,
\xxx9601003;

\bibitem{Douglas} M.R. Douglas, {\em Branes within Branes}, \xxx9512077;

\bibitem{ber} M. Bershadsky, C. Vafa, V. Sadov, {\em D-Branes and
Topological Field Theories}, \npb463(1996)420, \xxx9511222;

\bibitem{VW} C.Vafa and E. Witten, {\em A One Loop Test of String
Dualities}, \npb447(1995)261, \xxx9505053;

\bibitem{Hull} C.M. Hull and P.K. Townsend, {\em Unity of Superstring
Dualities}, \npb438(1995)109, \xxx9410167;

\bibitem{GHM} M.B. Green, J.A. Harvey and G. Moore, {\em I-Brane Inflow
and Anomalous Couplings on D-branes}, \cqg14(1997)47, \xxx9605033; 

\bibitem{bachas} C.P. Bachas, {\em Lectures on D-branes}, \xxx9806199;

\bibitem{CH} C.G. Callan, jr., and J.A. Harvey, {\em Anomalies
and Fermion Zero Modes on Strings and Domain Walls}, \npb250(1985)427;

\bibitem{CY} Y.-K.E. Cheung and Z. Yin, {\em Anomalies, Branes, and
Currents}, \npb517(1998)69, \xxx9710206;

\bibitem{MM} R. Minasian and G. Moore, {\em K-theory and Ramond-Ramond 
charge}, \jhep11(1997)002, \xxx9710230;

\bibitem{t1orb} A. Sagnotti, {\em Open Strings and their Symmetry
Groups}, Carg\`ese 1987, {\em Non-perturbative Quantum Field Theory},
eds. G. Mack {\em et al.}, Pergamon Press 1988; \\ P. Ho\v{r}ava, {\em
Background Duality of Open String Models}, \plb231(1989)251; \\ {\em
ibid.}, {\em Strings on Worldsheet Orbifolds}, \npb327(1989)461;

\bibitem{DLP} J. Dai, R.G. Leigh, J. Polchinski, {\em New Connections
Between String Theories}, \mpla4(1989)2073;

\bibitem{Gimon} E.G. Gimon, J. Polchinski, {\em Consistency Conditions
for Orbifolds and D-Manifolds}, \prd54(1996)1667, \xxx9601038;

\bibitem{Mukhor} S. Mukhi, {\em Orientifolds: The Unique Personality Of
Each Spacetime Dimension}, \xxx9710004;

\bibitem{PC} J. Polchinski and Y. Cai, {\em Consistency of Open
Superstring Theories}, \npb296(1988)91;

\bibitem{CLNY-1} C.G. Callan, jr., C. Lovelace, C.R. Nappi and S.A. Yost,
{\em Adding holes and Crosscaps to the Superstring}, \npb293(1987)83;

\bibitem{dasgupta} K. Dasgupta, D.P. Jatkar, S. Mukhi, {\em Gravitational 
Couplings and $Z_2$ Orientifolds}, \xxx9707224; \\ K. Dasgupta, S. Mukhi,
{\em Anomaly Inflow on Orientifold Planes}, \xxx9709219;

\bibitem{senm} A. Sen, {\em Strong Coupling Dynamics of Branes from
M-theory}, \jhep10(1997)002, \xxx9708002;

\bibitem{Craps} B. Craps and F. Roose, {\em Anomalous D-brane and 
Orientifold Couplings from the Boundary State}, \xxx9808074;

\bibitem{GS} M.B. Green and J.H. Schwarz, {\em Anomaly Cancellations in
Supersymmetric D=10 Gauge Theory and Superstring Theory},
\plb149(1984)117;

\bibitem{Green76} M.B. Green, {\em Reciprocal Spacetime and Momentum
Space Singularities in the Narrow Resonance Approximation},
\npb116(1976)449;

\bibitem{Green94} M.B. Green, {\em Point-Like States for Type IIB
Superstrings}, \plb329(1994)435, \xxx9403040;

\bibitem{cK} C.G. Callan, jr., and I.R. Klebanov, {\em D-brane Boundary
State Dynamics}, \npb465(1996)473, \xxx9511173;

\bibitem{Billo} M. Billo, P. Di Vecchia, M. Frau, A. Lerda, I. Pesando, R.
Russo, S. Sciuto, {\em Microscopic String Analysis of the D0-D8-Brane
System and Dual R-R States}, \xxx 9802088; 

\bibitem{diVecchia} P. Di Vecchia, M. Frau, I. Pesando, S. Sciuto, A.
Lerda, R. Russo, {\em Classical $p$-branes from Boundary State},
\xxx9707068; 

\bibitem{senbdr} A. Sen, {\em Stable Non-BPS States in String Theory},
\jhep06(1998)007, \xxx9803194;

\bibitem{GSei} M.B. Green and N. Seiberg, {\em Contact Interactions in
Superstring Theory}, \npb299(1988)559; 

\bibitem{Gutpole} M. Gutperle, {\em Contact Terms, Symmetries and
D-Instantons}, \xxx9705023; 

\bibitem{MSS} J.F. Morales, C.A. Scrucca and M. Serone {\em Anomalous
Couplings for D-branes and O-planes}, \xxx9812071;


\end{thebibliography}
\end{document}